\documentclass[
 reprint,
 superscriptaddress,
 amsmath,amssymb,
 aps,
 pra,
 nofootinbib
]{revtex4-1}

\usepackage{overpic}
\clearpage{}\usepackage[uline]{hhtensor}
\usepackage{mathrsfs}
\usepackage{hyperref}
\usepackage{mathtools}
\usepackage{xcolor}

\newcommand\duetoeq[1]{\stackrel{\text{Eq.\ (\ref{#1})}}{=}}

\newcommand\equaldueto[1]{\stackrel{#1}{=}}

\newcommand\Eq[1]{Eq.~(\ref{#1})}

\newcommand\rr{\mathbf{r}}
\newcommand\rrbar{\mathbf{\bar{r}}}
\newcommand\zhat{\mathbf{\hat{z}}}
\newcommand\xhat{\mathbf{\hat{x}}}
\newcommand\yhat{\mathbf{\hat{y}}}
\newcommand\pp{\mathbf{k}}

\newcommand\phat{\mathbf{\hat{\pp}}}

\newcommand\PP{\mathbf{P}}
\newcommand\JJ{\mathbf{J}}

\newcommand\Help{\ii\phat\times}

\newcommand\Gpplusstatic{\mathbf{F}_+(0,\pp)}
\newcommand\Gpplusstaticdagger{{\mathbf{F}_+(0,\pp)}^\dagger}
\newcommand\Gpminusstatic{\mathbf{F}_-(0,\pp)}
\newcommand\Gpminusstaticdagger{{\mathbf{F}_-(0,\pp)}^\dagger}

\newcommand\Gpplusdynamic{\mathbf{F}_+(\cz|\pp|,\pp)}
\newcommand\Gpplusdynamicdagger{{\mathbf{F}_+(\cz|\pp|,\pp)}^\dagger}
\newcommand\Gpminusdynamic{\mathbf{F}_-(\cz|\pp|,\pp)}
\newcommand\Gpminusdynamicdagger{{\mathbf{F}_-(\cz|\pp|,\pp)}^\dagger}

\newcommand\Gpplusminusdynamic{\mathbf{F}_\pm(\cz|\pp|,\pp)}
\newcommand\Gpplusminusstatic{\mathbf{F}_\pm(0,\pp)}

\newcommand\Ert{\mathbf{E}(t,\rr)}
\newcommand\Ethetart{\mathbf{E}_\theta(t,\rr)}
\newcommand\Drt{\mathbf{D}(t,\rr)}
\newcommand\Brt{\mathbf{B}(t,\rr)}
\newcommand\Bthetart{\mathbf{B}_\theta(t,\rr)}

\newcommand\Erealrt{\mathcal{E}(t,\rr)}

\newcommand\Brealrt{\mathcal{B}(t,\rr)}
\newcommand\Arealrt{\mathcal{A}(t,\rr)}
\newcommand\Crealrt{\mathcal{C}(t,\rr)}

\newcommand\Xrt{\mathbf{X}(t,\rr)}
\newcommand\Xomegapp{\mathbf{X}(\omega,\pp)}

\newcommand\intfourdwp{\int_{\omega\ge0}^\infty\text{ }\frac{d\omega}{\sqrt{2\pi}}\int_{\mathbb{R}^3} \frac{d \pp}{\sqrt{(2\pi)^3}}\text{ }}

\newcommand\intdpconf{\int_{\mathbb{R}^3} \frac{d \pp}{\cz|\pp|}\text{ }}

\newcommand\intdpnorm{\int_{\mathbb{R}^3} \frac{d \pp}{\sqrt{(2\pi)^3}}\text{ }}

\newcommand\intdr{\int_{\mathbb{R}^3} {d \rr}\text{ }}
\newcommand\intdrbar{\int_{\mathbb{R}^3} {d \bar{\rr}}\text{ }}

\newcommand\AVLambdatot{\langle \Lambda \rangle}
\newcommand\AVLambdafield{\langle \Lambda_{\text{field}} \rangle}
\newcommand\AVLambdastatic{\langle \Lambda_{\omega=0} \rangle}
\newcommand\AVLambdadynamic{\langle \Lambda_{\omega> 0} \rangle}

\newcommand\zerovec{\mathbf{0}}
\newcommand\mrest{\mathbf{M}(\rr)}

\newcommand\mrestpp{\mathbf{M}(0,\pp)}

\newcommand\mrestppplus{\mathbf{M}_+(0,\pp)}
\newcommand\mrestppminus{\mathbf{M}_-(0,\pp)}

\newcommand\mresttranspp{\mathbf{M}^{\perp}(0,\pp)}

\newcommand\rhorestpp{\rho(0,\pp)}
\newcommand\sigmarest{\Sigma(\rr)}
\newcommand\sigmart{\Sigma(t,\rr)}

\newcommand\ii{i}

\newcommand\jert{j_e(t,\rr)}
\newcommand\jmrt{j_m(t,\rr)}

\newcommand\rhoert{\rho_e(t,\rr)}
\newcommand\rhoerest{\rho_e(\rr)}
\newcommand\rhoerestbar{\rho_e(\rrbar)}
\newcommand\rhoerestpp{\rho_e(0,\pp)}
\newcommand\rhoethetart{\rho_e^\theta(t,\rr)}
\newcommand\rhomrt{\rho_m(t,\rr)}
\newcommand\rhomrest{\rho_m(\rr)}
\newcommand\rhomrestpp{\rho_m(0,\pp)}
\newcommand\rhomthetart{\rho_m^\theta(t,\rr)}

\newcommand\Jert{\mathbf{J}_e(t,\rr)}
\newcommand\Jer{\mathbf{J}_e(\rr)}
\newcommand\jerest{j_e(\rr)}
\newcommand\jmrest{j_m(\rr)}
\newcommand\Jethetart{\mathbf{J}_e^\theta(t,\rr)}

\newcommand\Jmrt{\mathbf{J}_m(t,\rr)}
\newcommand\Mrt{\mathbf{M}(t,\rr)}
\newcommand\Prt{\mathbf{P}(t,\rr)}

\newcommand\Jmthetart{\mathbf{J}_m^\theta(t,\rr)}

\newcommand\epsz{\epsilon_0}
\newcommand\muz{\mu_0}
\newcommand\cz{c_0}
\newcommand\Zz{Z_0}

\newcommand\Er{\mathbf{E}(\rr)}
\newcommand\Ep{\mathbf{E}(0,\pp)}
\newcommand\Epdyn{\mathbf{E}(\cz|\pp|,\pp)}
\newcommand\Br{\mathbf{B}(\rr)}

\newcommand\Ar{\mathbf{A}(\rr)}

\newcommand\Bp{\mathbf{B}(0,\pp)}
\newcommand\Bpdyn{\mathbf{B}(\cz|\pp|,\pp)}
\newcommand\Bwp{\mathbf{B}(\omega,\pp)}

\newcommand\Ewp{\mathbf{E}(\omega,\pp)}

\newcommand\Bpplus{\mathbf{B}_+(0,\pp)}
\newcommand\Bpminus{\mathbf{B}_-(0,\pp)}
\newcommand\Ap{\mathbf{A}(0,\pp)}
\newcommand\Aptrans{\mathbf{A}^{\perp}(0,\pp)}
\newcommand\Bpdagger{{\mathbf{B}(0,\pp)}^\dagger}
\newcommand\Gplambda{\mathbf{F}_\pm(0,\pp)}
\newcommand\Gplambdadagger{{\mathbf{F}_\pm(0,\pp)}^\dagger}

\clearpage{}

\begin{document}
\title{The total helicity of electromagnetic fields and matter}
\author{Ivan Fernandez-Corbaton}
\affiliation{Institute of Nanotechnology, Karlsruhe Institute of Technology, 76021 Karlsruhe, Germany}
\email{ivan.fernandez-corbaton@kit.edu}
\begin{abstract}
	The electromagnetic helicity of the free electromagnetic field and the static magnetic helicity are shown to be two different embodiments of the same physical quantity, the total helicity. The total helicity is the sum of two terms: A term that measures the difference between the number of left-handed and right-handed photons of the free field, and another term that measures the screwiness of the static magnetization density in matter. Each term is the manifestation of the total helicity in different frequency regimes: $\omega>0$ and $\omega=0$, respectively. This unification establishes the theoretical basis for studying the conversion between the two embodiments of total helicity upon light-matter interaction.     
\end{abstract}
\maketitle
\section{Introduction and summary}
The electromagnetic helicity \cite{Calkin1965,Zwanziger1968,Deser1976,Birula1981,Birula1996,Afanasiev1996,Trueba1996,Drummond1999} is a property of the free electromagnetic field that extends the concept of circular polarization handedness from individual plane waves to general Maxwell fields. Its integrated value is a pseudo-scalar proportional to the difference between the number of left- and right- handed photons contained in the field. A recently renewed interest in electromagnetic helicity \cite{Coles2012,FerCor2012p,Cameron2012,Bliokh2013,Cameron2013,FerCorTHESIS,Nieto2015,Gutsche2016,FerCor2016,Elbistan2017,Andrews2018,Vazquez2018,Hanifeh2020,Crimin2019,FernandezGuasti2019,Poulikakos2019,Bernabeu2019} is revealing and exploiting its effectiveness for understanding and engineering light-matter interactions, in particular at the challenging micro and nano scales. Such effectiveness is, to some extent, due to the connection between electromagnetic helicity and the electromagnetic duality symmetry of free fields \cite{Calkin1965,Zwanziger1968}, which greatly facilitates the use of symmetries and conservation laws in the analysis of light-matter interactions. Electromagnetic helicity is, in many ways, at the same level of generality as electromagnetic energy, momentum and angular momentum: It is a measurable property of the field which is connected to a fundamental symmetry transformation. But there is an important difference: While energy, momentum, and angular momentum are also defined for material systems, and the possibility and effects of the exchange of such properties between fields and matter are theoretically understood and practically exploited, the same is not true for electromagnetic helicity. It is so far unclear whether a material system can have electromagnetic helicity, which means that we lack the theoretical basis for considering an exchange of electromagnetic helicity between fields and matter. This is an unsatisfactory state of affairs, in particular because the integrated electromagnetic helicity of the field is typically different before and after a light-matter interaction.

In this article, we obtain the definition of the total helicity of fields and matter by a natural extension of the definition of the integrated electromagnetic helicity of the free field. The complete definition contains an additive contribution from static material sources. Such contribution turns out to be the magnetic helicity \cite{Woltjer1958,Moffatt1969,Ranada1992}. The resulting total helicity is equal to the sum of the screwiness of the static magnetization density in matter plus the difference between the number of left- and right- handed photons in the free electromagnetic field. The unification provides the theoretical basis for studying the conversion between the two embodiments of total helicity upon light-matter interaction. In particular, our result implies that material systems able to sustain static magnetization configurations with some degree of screwiness, have the potential of storing electromagnetic helicity coming from the dynamic electromagnetic free field, and of returning the stored helicity to the free field by means of electromagnetic radiation.

We highlight that, in this article, we do not study the possibility of a conservation law for the electromagnetic helicity including both fields and sources. The questions that we are rather addressing aim at establishing whether the exchange of electromagnetic helicity between fields and sources is at all meaningful. A positive answer is apparently a pre-requisite for considering a joint light-matter conservation law for electromagnetic helicity.

The rest of the article is organized as follows. In Sec.~\ref{sec:question}, we consider the question {\em can the electromagnetic free field and a material system exchange electromagnetic helicity?}, which is motivated by the fact that the integrated electromagnetic helicity of the field is typically different before and after the light-matter interaction. The question forces upon us the need for identifying, for a material system, the counterpart of the electromagnetic helicity for the free field. In Sec.~\ref{sec:withsources}, we follow the example of electrostatic energy and consider the static electromagnetic sources or, alternatively and equivalently, the static fields that they generate, as the potential reservoirs of electromagnetic helicity. More precisely, we consider sources in static equilibrium where the time derivatives of macroscopic quantities vanish. Under such conditions, we show that the most commonly assumed Maxwell sources cannot act as a reservoir of helicity. This roadblock is then removed by assuming electric charge and magnetic spin as the primordial sources, instead of electric charge and magnetic charge, or electric charge only. Then, we argue in Sec.~\ref{sec:total} that the definition of the integrated electromagnetic helicity of the free field is incomplete, in the sense that it does not include the whole domain of definition of electromagnetic fields. Its natural completion includes a static contribution coming from the transverse(divergence-free) part of the static spin magnetization density, and results in the definition of the total helicity. The static contribution turns out to be the magnetic helicity, albeit in different units. The electromagnetic helicity of the free field, and the static magnetic helicity, are hence manifestations of the total helicity in different frequency regimes: $\omega>0$ and $\omega=0$, respectively. Section~\ref{sec:conclusion} contains concluding remarks and a brief indication of the potential impact of the findings.

\section{Motivation and problem setting\label{sec:question}}
For a given free electromagnetic field, the integrated value of the electromagnetic helicity is a pseudo-scalar, proportional to the difference between the number of left- and right-handed polarized photons contained in the field. For the free field, this pseudo-scalar is a constant of the time evolution because of the electromagnetic duality symmetry. That is, because the equations
\begin{equation}
	\label{eq:menosources}
	\begin{split}
		&\nabla\cdot\Brt=0,\ \cz^2\nabla\times\left[\epsz\Ert\right]+\frac{\partial_t \Brt}{\muz}=\zerovec,\\
		&\nabla\cdot\Ert=0,\ \nabla\times\Brt-\frac{\partial_t\Ert}{\cz^2}=\zerovec,
	\end{split}
\end{equation}
are invariant under the duality transformation\footnote{The name ``electromagnetic duality'' is also often used to refer to a discrete version, recovered from \Eq{eq:d} when $\pi=\pi/2$: $\mathbf{E}_{\pi/2}(\rr,t)=-\cz\Brt$, $\cz\mathbf{B}_{\pi/2}(\rr,t)=\Ert$.}
\begin{equation}
	\label{eq:dnosources}
	\begin{split}
		\Ethetart&=\Ert\cos\theta-\cz\Brt\sin\theta,\\
		\cz\Bthetart&=\Ert\sin\theta+\cz\Brt\cos\theta,\\
	\end{split}
\end{equation}
where $\theta$ is a real angle. SI units will be used throughout the article.
\begin{figure}
	\includegraphics[width=\linewidth]{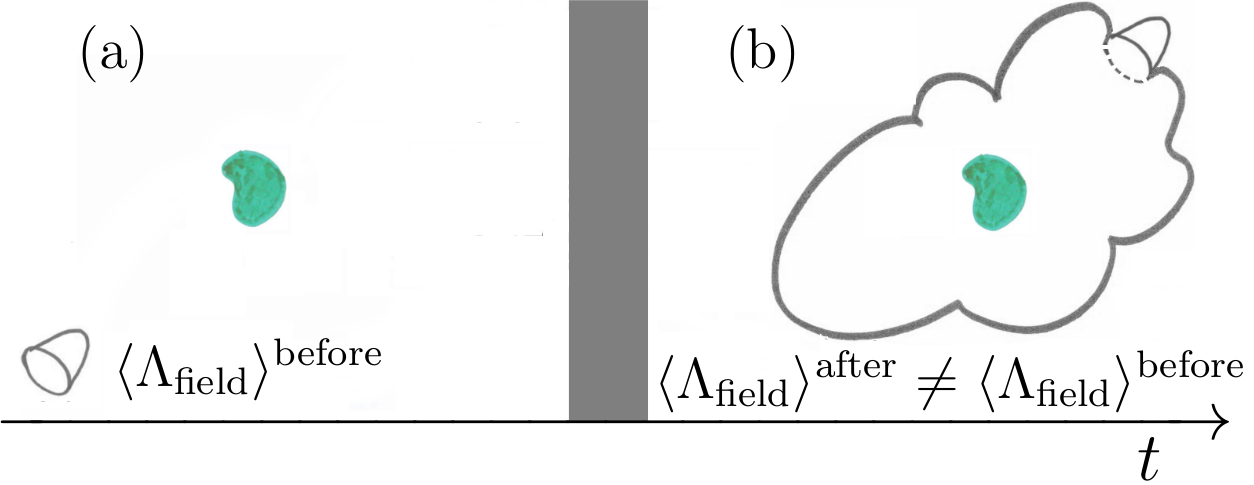}
	\caption{ \label{fig:interaction} Sequential phases of the light-matter interaction. The material system is represented by the green objects. The incoming bullet-shaped field in (a), and the outgoing cloud-plus-bullet-shaped field in (b) do not interact with the material system. The interaction occurs during the grayed-out region.  The integrated electromagnetic helicity of the field $\AVLambdafield$ is typically different before and after the interaction. We address the question of whether a part of that difference can be stored in the material system.}
\end{figure}
	The invariance under duality transformations is typically lost once electromagnetic sources are involved. Then, the integrated electromagnetic helicity of the field is typically different before and after the light-matter interaction. Let us consider the light-matter interaction time sequence depicted in Fig.~\ref{fig:interaction}. Before the interaction, in panel (a), an incoming field (gray bullet) approaches the material system (green object). The field and the material system interact during the gray period. After the interaction, in panel (b), the resulting outgoing beam propagates away from the material system. During the interaction, the field and the matter can exchange different measurable properties, having different consequences for the material system. For example, when the field transfers energy to the material system, the system may undergo a transition to an excited state. When the field transfers linear(angular) momentum to the material system, the system will experience a force(torque). Electromagnetic energy, momentum, and angular momentum are defined for both fields and matter, and the exchanges of these quantities during light-matter interaction can be seen as a conversion between two different embodiments of the same fundamental property. The situation regarding electromagnetic helicity is much less clear. While there is no question that the integrated electromagnetic helicity of the free field can be different before and after the light-matter interaction ($\AVLambdafield^{\text{after}}\neq\AVLambdafield^{\text{before}}$), the existence of such a thing as the electromagnetic helicity of a material system, where part of that difference can end, up is unclear. Then, the well-posedness of questions like {\em Can the electromagnetic free field and a material system exchange electromagnetic helicity?} or {\em What happens to the material system when the field transfers electromagnetic helicity to it?} is also unclear. Before progress can be made regarding such questions, we must elucidate whether there exists a way to store helicity in matter as just a different form of the same physical quantity as the electromagnetic helicity of the free field. Such is the case, for example, for energy, which can be contained in the free field, and also stored in a static electric charge density distribution in the form of electrostatic energy \cite[Eqs.~(1.17,4.83,4.89)]{Jackson1998}: 
	\begin{equation}
		\label{eq:we}
		W_{e}=\frac{\epsilon_0}{2}\intdr \Er\cdot\Er=\frac{1}{8\pi\epsilon_0}\intdr\intdrbar\frac{\rhoerest\rhoerestbar}{|\rr-\rrbar|}.
	\end{equation}
	We can interpret that the energy is either stored in the source $\rhoerest$, or in the $\Er$ field produced by it. The two interpretations are equivalent due to the one-to-one relationship between fields and sources in the static case. This leads us to consider the static electromagnetic sources or, alternatively and equivalently, the fields produced by them, as the potential reservoirs of electromagnetic helicity in matter. Consequently, in the next section we will consider the static version of Maxwell equations with sources.

We highlight that the setting illustrated in Fig.~\ref{fig:interaction} provides a clear distinction between the dynamic free fields and the static fields that are inherent to matter. This distinction avoids possible ambiguities that arise when effective macroscopic dynamic fields inside the material are considered, which, in this setting, would only exist during the gray period of Fig.~\ref{fig:interaction}. Such effective fields have a mixed field-matter character which complicates the aforementioned distinction. The century-old Abraham-Minkowski controversy regarding the split of the linear momentum in dielectric media between fields and matter is a prominent example of such complication.

\section{Electric charge and magnetic spin as primordial Maxwell sources\label{sec:withsources}}
	The study of electromagnetic helicity in the presence of fundamental sources has previously been based in the following microscopic Maxwell equations, either including \cite{Zwanziger1968} or excluding \cite{Nieto2015,Nienhuis2016} the magnetic sources:
	 {\footnotesize
\begin{equation}
	\label{eq:me}
	\begin{split}
		&\nabla\cdot\Brt=\muz\rhomrt,\ -\cz^2\nabla\times\left[\epsz\Ert\right]=\frac{\partial_t \Brt}{\muz}+\Jmrt,\\
		&\nabla\cdot\Ert=\frac{\rhoert}{\epsz},\ \nabla\times\Brt=\muz\left[\Jert+\epsz\partial_t\Ert\right],
	\end{split}
\end{equation}
}
	where $\rho_{e/m}(\rr,t)$ and $\mathbf{J}_{e/m}(\rr,t)$ are the electric/magnetic charge and current density distributions, respectively. In particular, Zwanziger \cite{Zwanziger1968} used \Eq{eq:me} to show that electrodynamics is invariant when, besides applying \Eq{eq:dnosources} to the fields, the duality transformation is also applied to the sources:
\begin{equation}
	\label{eq:d}
	\begin{split}
		\rhoethetart&=\rhoert\cos\theta-\cz\rhomrt\sin\theta,\\
		\cz\rhomthetart&=\rhoert\sin\theta+\cz\rhomrt\cos\theta,\\
		\Jethetart&=\Jert\cos\theta-\cz\Jmrt\sin\theta,\\
		\cz\Jmthetart&=\Jert\sin\theta+\cz\Jmrt\cos\theta.\\
	\end{split}
\end{equation}
	 Let us examine the static version of Eqs.~(\ref{eq:me}). To such end, it is important to examine the assumptions underlying Eqs.~(\ref{eq:me},\ref{eq:d}). Namely, that there exist primordial electric and magnetic elementary charges which result in electric and magnetic charge densities inside material systems, and that {\em the electric and magnetic current densities are due to the movement of the electric and magnetic charge densities, respectively}. The charge and current densities transform together as four-vectors under the Poincar\'e group of special relativity:  
	\begin{equation}
		\jert=\begin{bmatrix}\rhoert\\\Jert\end{bmatrix},\ \jmrt=\begin{bmatrix}\rhomrt\\\Jmrt\end{bmatrix}.
	\end{equation}
	The static sources are hence
	\begin{equation}
		\label{eq:jejmrest}
		\jerest=\begin{bmatrix}\rhoerest\\\zerovec\end{bmatrix},\ \jmrest=\begin{bmatrix}\rhomrest\\\zerovec\end{bmatrix},
	\end{equation}
	where the vanishing of the 3-vector currents $\mathbf{J}_{e/m}(\rr)=\zerovec$ is due to the vanishing of net macroscopic movement of the static charge densities $\rhoerest$ and $\rhomrest$ inside the material system. While there will generally be some microscopical dynamics, like e.g. due to thermal fluctuations, we will assume that the material system before the light-matter interaction in Fig.~\ref{fig:interaction}(a), and after it in Fig.~\ref{fig:interaction}(b) is in a state of static equilibrium where the time derivatives of macroscopic quantities vanish. This includes the vanishing of the macroscopic time derivative of the position of the electric and magnetic charge densities, and hence the vanishing of $\mathbf{J}_{e/m}(\rr)$. The static equilibrium version of \Eq{eq:me} is hence obtained by eliminating all the terms containing time derivatives, and using the sources from \Eq{eq:jejmrest}:
\begin{equation}
	\label{eq:mezero}
	\begin{split}
		\nabla\cdot\Er=\frac{\rhoerest}{\epsz}&,\ \nabla\cdot\Br=\muz\rhomrest,\\
		\cz^2\nabla\times\left[\epsz\Er\right]=\zerovec,&\ \nabla\times\Br=\zerovec.
	\end{split}
\end{equation}

\begin{table}[t!]
\begin{center}\begin{tabular}{c c c c c} 
	$(t,\rr):$&$\Xrt$&$\partial_t\Xrt$&$\nabla\cdot\Xrt$&$\nabla\times\Xrt$\\
	&$\updownarrow$&$\updownarrow$&$\updownarrow$&$\updownarrow$\\
	$(\omega,\pp):$&$\Xomegapp$&$-\ii\omega\Xomegapp$&$\ii\pp\cdot\Xomegapp$&$\ii\pp\times\Xomegapp$\\
 \end{tabular} 
	\end{center}
 \caption{Correspondences between operators in $(t,\rr)$ and operators in $(\omega,\pp)$.}
\label{tab:xtop}
\end{table}

	According to our previous discussion, the electromagnetic helicity of the material system before and after the light-matter interaction is contained in the configuration of the static sources in panels (a) and (b), respectively, or, equivalently, in the static fields produced by such static sources. We now show that the static electromagnetic fields in \Eq{eq:mezero} cannot store helicity because both $\Er$ and $\Br$ have vanishing curl (are longitudinal). To such end, we consider the Fourier-transformed version of \Eq{eq:mezero}, which is obtained using the correspondences between the time-space $(t,\rr)$-domain and the frequency-wavevector $(\omega,\pp)$-domain contained in Tab.~\ref{tab:xtop} for $\Xomegapp$ functions, particularized for the time-independent $\omega=0$ case:
\begin{equation}
	\label{eq:mezeropp}
	\begin{split}
		\ii\pp\cdot\Ep=\frac{\rhoerestpp}{\epsz}&,\ \ii\pp\cdot\Bp=\muz\rhomrestpp,\\
		-\cz^2\ii\pp\times\left[\epsz\Ep\right]=\zerovec,&\ \ii\pp\times\Bp=\zerovec.
	\end{split}
\end{equation}
	The two $\ii\pp\times$ equations in \Eq{eq:mezeropp} imply that both $\mathbf{E}(0,\pp)$ and $\mathbf{B}(0,\pp)$ are purely longitudinal, that is, $\Ep$ and $\Bp$ are parallel to the wavevector $\pp$, having zero components that are transverse (perpendicular) to $\pp$. Purely longitudinal(transverse) fields in the $\pp$-domain correspond to fields with zero curl(divergence) in the $\rr$-domain. Let us now consider the $\pp$-domain representation of the helicity operator, which follows\footnote{$\Lambda=\frac{\mathbf{J}\cdot\mathbf{P}}{|\mathbf{P}|}=\frac{\mathbf{S}\cdot\mathbf{P}}{|\mathbf{P}|}\equiv\Help$, where for electromagnetism, $\mathbf{S}$ is the vector of spin-1 matrices. The second equality in the previous equation can be seen to follow, for example, from considering the coordinate representation of the angular momentum and linear momentum operator vectors, \cite[Eqs.~(5.24,5.25)]{Birula1996}: $\mathbf{J}\equiv -\ii\rr\times\nabla + \mathbf{S}$, $\mathbf{P}\equiv -i\nabla$. Their inner product then reads $\mathbf{J}\cdot\mathbf{P}\equiv -(\rr\times\nabla)\cdot\nabla-i \mathbf{S}\cdot\nabla$. The first term vanishes since it is the divergence of a curl. Finally, the equivalence $\frac{\mathbf{S}\cdot\mathbf{P}}{|\mathbf{P}|}\equiv\Help$ follows from applying \cite[Eq. (2.2)]{Birula1996} in wavevector space where $\mathbf{P}\rightarrow \pp\implies\mathbf{P}/|\mathbf{P}|\rightarrow \phat$. We have assumed $\hbar=1$.} from the general definition of helicity as the projection of the angular momentum operator ($\JJ$) onto the linear momentum ($\PP$) direction:
\begin{equation}
	\label{eq:help}
	\Lambda=\frac{\mathbf{J}\cdot\mathbf{P}}{|\mathbf{P}|}\equiv\frac{\ii\pp\times}{|\pp|}=\Help.
\end{equation}

The reduced Planck constant $\hbar$ is implicitly set to 1 in \Eq{eq:help}, and for the rest of the paper. Without this assumption, the helicity operator reads $\hbar\Help$, and its three eigenvalues are $\pm\hbar$ and $0$. 

	Applying the helicity operator to the static electric(magnetic) fields from \Eq{eq:mezeropp} results in the zero field: 
	\begin{equation}
		\Help\Ep=\Help\Bp\duetoeq{eq:mezeropp}\zerovec,\\
	\end{equation}
	showing that the static electromagnetic fields in \Eq{eq:mezeropp} cannot store any electromagnetic helicity. The conclusion is the same if the magnetic sources in \Eq{eq:me} are removed \cite{Nienhuis2016}, and the vanishing of $\mathbf{J}_e(\rr)$ is maintained. In such case, $\Ep$ is longitudinal and $\Bp=\zerovec$.

We have reached the conclusion that the model underlying \Eq{eq:me} implies that material systems in static equilibrium cannot store electromagnetic helicity. This issue must be added to the lack of experimental evidence for isolated magnetic charges (magnetic monopoles).

It turns out that the two issues can be overcome by adopting a different set of primordial electromagnetic sources for Maxwell equations. Supported by the fact that the existence of magnetic spin is beyond doubt, we will now consider electric charges and magnetic spins as the primordial sources, instead of electric and magnetic charges, or electric charges only. The static equilibrium sources that we assume from now on are
	\begin{equation}
		\label{eq:rhosigma}
		\jerest=\begin{bmatrix}\rhoerest\\\zerovec\end{bmatrix}, \sigmarest=\begin{bmatrix}0&0&0&0\\0&0&-M_3(\rr)&M_2(\rr)\\0&M_3(\rr)&0&-M_1(\rr)\\0&-M_2(\rr)&M_1(\rr)&0\end{bmatrix},
	\end{equation}
	where $\jerest$ transforms as a four-vector, and $\sigmarest$ is an antisymmetric tensor which transforms like the electromagnetic tensor\footnote{$\cz F\equiv\begin{bmatrix}0&-E_1(t,\rr)&-E_2(t,\rr)&-E_3(t,\rr)\\E_1(t,\rr)&0&-\cz B_3(t,\rr)&\cz B_2(t,\rr)\\E_2(t,\rr)&\cz B_3(t,\rr)&0&-\cz B_1(t,\rr)\\E_3(t,\rr)&-\cz B_2(t,\rr)&\cz B_1(t,\rr)&0\end{bmatrix}$.} $F$. The three distinct components of $\sigmarest$ constitute the static spin magnetization density $\mrest$. The spatial integral of $\mrest$ over the volume of the material system defines the intrinsic magnetic moment of the system in static equilibrium. With these assumptions, and for our purposes, the question of whether to model magnetic effects by microscopic electric current loops or microscopic magnetic dipoles \cite[Chap.~2, \S~1]{Brown1962} is decided in favor of the latter. 
	
	The movement of charge and spin results in the dynamic sources
	\begin{equation}
		\label{eq:rhosigmart}
		\begin{split}
			\jert&=\begin{bmatrix}\rhoert\\\Jert\end{bmatrix},\\
				\sigmart&=\begin{bmatrix}0&-\cz P_1(t,\rr)&-\cz P_2(t,\rr)&-\cz P_3(t,\rr)\\\cz P_1(t,\rr)&0&-M_3(t,\rr)&M_2(t,\rr)\\\cz P_2(t,\rr)&M_3(t,\rr)&0&-M_1(t,\rr)\\\cz P_3(t,\rr)&-M_2(t,\rr)&M_1(t,\rr)&0\end{bmatrix},
		\end{split}
	\end{equation}
where, as before, $\Jert$ appears due to the movement of $\rhoert$. Additionally, the movement of $\sigmarest$ produces a dynamic $\sigmart$ which contains both magnetic spin density $\Mrt$, and electric spin density $\Prt$. Note that $\Prt$ is a source of electric kind originating from the magnetic spin, as opposed to the components of $\jert$, which originate from electric charge. Accordingly, the two kinds of electric sources have different transformation properties under the Poincar\'e group. In particular, $\Prt$ should not be confused with the dipolar density that can result from the action of an external field on the electric charge density.

In here, we use Eqs.~(\ref{eq:rhosigma},\ref{eq:rhosigmart}) for an extended material system. Their point-particle versions have a long history in the study of relativistic electrodynamics \cite[Chap.~II, Sec.~4]{Barut1980}, including the effect of the electron spin on the atomic nucleus \cite{Frenkel1926}, and the relativistic spin precession \cite{Bargmann1959}. In that context, the spatial integral of $\sigmart$ is often called dipole moment tensor, moment tensor, or  polarization tensor.

The sources in \Eq{eq:rhosigmart} result in a version of Maxwell's equations \cite[Sec.~5]{Holten1991} that is quite different from \Eq{eq:me}:
\begin{equation}
	\label{eq:hom}
	\nabla\cdot\Brt=\zerovec,\ \nabla\times\Ert+\partial_t \Br=\zerovec, \text{ and}
\end{equation}
\begin{equation}
	\label{eq:inhom}
	\begin{split}
		&\nabla\cdot\Ert=\frac{\rhoert-\nabla\cdot\Prt}{\epsz},\\ 
		&\cz^2\nabla\times\Brt-\partial_t\Ert=\\&
		\frac{1}{\epsz}\left[\Jert+\partial_t\Prt+\nabla\times\Mrt\right],
	\end{split}
\end{equation}
where the magnetic sources are of a different kind, and appear in a different position with respect to \Eq{eq:me}. In particular, the homogeneous equations in \Eq{eq:hom} contain the statement that there are no magnetic monopoles, and the divergence of $\Brt$ is always zero, making it a purely transverse field in all cases. This difference is crucial for enabling static sources to store electromagnetic helicity.

Let us now consider the static equilibrium limit of Eqs.~(\ref{eq:hom},\ref{eq:inhom}). Noting that both $\Jert$ and $\Prt$ vanish, we obtain:
\begin{equation}
	\label{eq:staticeqs}
	\begin{split}
		\nabla\cdot\Er=\frac{\rhoerest}{\epsilon_0}&,\ \nabla\times\Er=\zerovec,\\
		\nabla\cdot\Br=\zerovec&,\ \nabla\times\Br=\muz\nabla\times\mrest,\\
	\end{split}
\end{equation}

The first line in \Eq{eq:staticeqs} are the common equations that define the electrostatic field $\Er$ \cite[Chap.~4]{Jackson1998}. The second line in \Eq{eq:staticeqs} coincides with the common equations that define the magnetostatic field $\Br$, if, according to our previous discussion, the electric current density $\Jer$ that appears in those common equations (see e.g. \cite[Eq.~(2.40)]{Brown1962}, or \cite[Eqs.~(5.80,5.82)]{Jackson1998}) vanishes.

 \begin{table*}[t]
\begin{center}\begin{tabular}{c} 
	$\Xomegapp=\mathbf{X}^\parallel(\omega,\pp)+\mathbf{X}^\perp(\omega,\pp)$\\
	
	\\
	$\mathbf{X}^\parallel(\omega,\pp)=\phat\left[\phat\cdot\mathbf{X}(\omega,\pp)\right]$\\

	\\
	${\mathbf{X}^\parallel(\omega,\pp)}^\dagger\mathbf{X}^\perp(\omega,\pp)=0$\\	
	\\
	$\Help\mathbf{X}_\lambda(\omega,\pp)=\lambda\mathbf{X}_\lambda(\omega,\pp)$ for $\lambda\in[-1,0,+1]\text{ ($\hbar=1$ is assumed)}$\\
	\\
	$\mathbf{X}_\lambda(\omega,\pp)=x_\lambda(\omega,\pp)\mathbf{\hat{e}}_\lambda(\phat)$\\

	\\
	$\mathbf{X}(\omega,\pp)=\mathbf{X}_-(\omega,\pp)+\mathbf{X}_0(\omega,\pp)+\mathbf{X}_+(\omega,\pp)$\\ 
	\\
		$\mathbf{X}_0(\omega,\pp)=\mathbf{X}^\parallel(\omega,\pp)$, $\mathbf{X}^\perp(\omega,\pp)=\mathbf{X}_+(\omega,\pp)+\mathbf{X}_-(\omega,\pp)$\\

	\\
	${\mathbf{X}_0(\omega,\pp)}^\dagger\mathbf{X}_+(\omega,\pp)={\mathbf{X}_0(\omega,\pp)}^\dagger\mathbf{X}_-(\omega,\pp)={\mathbf{X}_+(\omega,\pp)}^\dagger\mathbf{X}_-(\omega,\pp)=0$\\

	\\
	$\left(\Help\right)^2\Xomegapp=\Help\Help\Xomegapp=\mathbf{X}^\perp(\omega,\pp)=\mathbf{X}(\omega,\pp)-\mathbf{X}^\parallel(\omega,\pp)$\\
 \end{tabular} 
	\end{center}
 \caption{Various identities involving the decomposition of a vectorial $\Xomegapp$ function in terms of its longitudinal $(\parallel)$ and transverse parts $(\perp)$ (Helmholtz decomposition), and also in terms of the eigenvectors of the helicity operator $\Lambda$ corresponding to its three eigenvalues $[-1,0,+1]$ ($\hbar=1$ is assumed). The symbol $^\dagger$ denotes complex transposition, $x_\lambda(\omega,\pp)$ are complex scalar functions, $\phat=\pp/|\pp|$, and $\mathbf{\hat{e}}_\lambda(\phat)$ are the $\phat$-dependent unit vectors which can be obtained by rotating those corresponding to $\phat=\zhat$ (see e.g. \cite[Eq.~(8.7-11)]{Tung1985}): $\mathbf{\hat{e}}_\lambda(\phat)=R_z(\phi)R_y(\beta)\mathbf{\hat{e}}_\lambda(\zhat)$, where $\phi=\arctan(p_y/p_x), \beta=\arccos(p_z/|\pp|)$, and $\mathbf{\hat{e}}_0(\zhat)=\zhat$, $\sqrt{2}\mathbf{\hat{e}}_\pm(\zhat)=\mp\xhat-\ii\yhat$.}
\label{tab:decomphel}
\end{table*}

\section{The total helicity of fields and matter\label{sec:total}}
We now proceed to show that, when the fundamental static sources from \Eq{eq:rhosigma} are assumed, the typical definition of integrated dynamic electromagnetic helicity can be naturally completed to include a static contribution. This contribution turns to be the magnetic helicity \cite{Woltjer1958,Moffatt1969,Ranada1992}, albeit in different units.

It is now convenient to change from the $(t,\rr)$-domain to the $(\omega,\pp)$-domain by means of the 4D-Fourier decomposition 
{\small
\begin{equation}
	\label{eq:4dft}
	\Xrt=\intfourdwp\Xomegapp\exp\left(-\ii\omega t+\ii\pp\cdot\rr\right),
\end{equation}
}
where only frequencies $\omega\ge 0$ are included: $\omega=0$ corresponds to the static fields and $\omega>0$ to the dynamic fields. Excluding $\omega<0$ amounts to considering complex dynamic fields with only positive energy. This is possible in electromagnetism because both sides of the spectrum contain the same information \cite[\S 3.1]{Birula1996}\cite{Birula1981}, and only one sign of the frequency(energy) is needed. 

In the following, we will often use properties of the decomposition of a $\Xomegapp$ function in terms of its longitudinal ($\parallel$) and transverse ($\perp$) parts, and in terms of the eigenvectors of the helicity operator $\Lambda$, which are collected in Tab.~\ref{tab:decomphel}, and the correspondences between operators in $(t,\rr)$ and operators in $(\omega,\pp)$ collected in Tab.~\ref{tab:xtop}.

The integrated electromagnetic helicity of the dynamic fields $\AVLambdadynamic$ (called $\AVLambdafield$ before, and in Fig.~\ref{fig:interaction}) can be computed in the $(\omega,\pp)$-domain as \cite{FerCor2019b}:
\begin{equation}
	\label{eq:Hel}
	\begin{split}
	&\AVLambdafield=\AVLambdadynamic=\\
			&\intdpconf\Gpplusdynamicdagger\Help\Gpplusdynamic\\
			&+\Gpminusdynamicdagger\Help\Gpminusdynamic,
	\end{split}
\end{equation}
where the symbol $^{\dagger}$ denotes complex transposition. The $\Gpplusminusdynamic$ are the plane wave components of a version of the Riemann-Silberstein vectors \cite{Birula1996,Birula2013} 
\begin{equation}
	\label{eq:gp}
	\begin{split}
	&	\frac{\Drt}{\sqrt{2\epsilon_0}} \pm i\frac{\Brt}{\sqrt{2\mu_0}}= \sqrt{\frac{\epsilon_0}{2}}\left[\Ert\pm i\cz\Brt\right]=\mathbf{F}_{\pm}(t,\rr)\\&=\intdpnorm \Gpplusminusdynamic\exp(i\pp\cdot\rr-i\cz|\pp| t),
	\end{split}
\end{equation}
where $\omega$ is restricted to be equal to $\cz|\pp|$ because, for $\omega>0$, the dynamic electromagnetic fields $\left[\Ewp,\Bwp\right]$ are constrained to the domain $\omega=\cz|\pp|$. This is the well-known constraint to the positive energy light-cone, which may be seen as a consequence of the massless photonic dispersion relations in vacuum $\omega^2=\cz^2\pp\cdot\pp$, together with the $\omega>0$ choice.

We note that, should we not have set $\hbar=1$, the factor $\frac{d\pp}{\cz|\pp|}$ in \Eq{eq:Hel} would instead read $\frac{d\pp}{\cz|\pp|\hbar}$. Recalling that the helicity operator would then be $\hbar\Help$, we see that the expression inside the integral in \Eq{eq:Hel} is independent of $\hbar$. Dimensional analysis shows that $\AVLambdafield$ has units of angular momentum, which are the units of helicity.

The time-dependent electromagnetic fields $\mathbf{F}_{\pm}(t,\rr)$ from before and after the light-matter interaction in Fig.~\ref{fig:interaction} can be exactly recovered from their $(\omega=\cz|\pp|>0,\pp)$-domain counterparts $\Gpplusminusdynamic$ by means of \Eq{eq:gp}. For the ``before'' electromagnetic field in Fig.~\ref{fig:interaction}(a), the result of the integral in \Eq{eq:gp} will be valid for all times prior to the start of the interaction. Similarly, for the ``after'' electromagnetic field in Fig.~\ref{fig:interaction}(b), the result of the integral in \Eq{eq:gp} will be valid for all times after the end of the interaction. Then, in the same way that the integrated energy or momentum of a free non-interacting field are constant in time\footnote{For free, non-interacting fields, the spatio-temporal evolution of $\mathbf{F}_{\pm}(t,\rr)$ is only due to the exponential in \Eq{eq:gp}, while the $\Gpplusminusdynamic$ components do not change with time. Then, the fact that the integrated values of measurable properties can be written as $\pp$-space integrals involving the $\Gpplusminusdynamic$ (see e.g. \cite[Eq.~(4.13)-(4.16)]{Birula1996}, and \cite[Eq.~(10)]{FerCor2019}), shows that the value of such integrals does not change with time.}, the integrated electromagnetic helicities of the ``before'' and ``after'' fields will be constant, albeit typically different from each other, in each of the time periods when the fields are defined.

Reference~\onlinecite{FerCor2019b} contains the proof of the equivalence between \Eq{eq:Hel} and the most common expression for integrated electromagnetic helicity \cite{Calkin1965,Zwanziger1968,Deser1976,Afanasiev1996,Trueba1996,Drummond1999,Cameron2012,Bliokh2013,Cameron2013,Nieto2015,Gutsche2016,Elbistan2017,Crimin2019,FernandezGuasti2019,Poulikakos2019,Bernabeu2019}, which reads:
\begin{equation}
	\label{eq:avhelreal}
	\AVLambdadynamic=\frac{1}{2}\intdr \Brealrt\cdot\Arealrt-\Erealrt\cdot\Crealrt,
\end{equation}
where $\Erealrt\left[\Crealrt\right]$ and $\Brealrt\left[\Arealrt\right]$ are the real-valued electric and magnetic fields[potentials], respectively.

We now want to understand whether there exists a more general quantity that comprises the electromagnetic helicity of the free field, and a contribution from matter. Should such more general quantity exist, we also want to write down its definition. We proceed by considering the fact that the complete $(\omega,\pp)$-domain of electric and magnetic fields is $(\omega=\cz|\pp|>0,\pp)$, which corresponds to dynamic fields, and $(\omega=0,\pp)$, which corresponds to static fields. Crucially, the action of the helicity operator $\Help$ is well-defined in both dynamic and static cases. These facts suggest that \Eq{eq:Hel} is incomplete in the sense that it does not cover the whole domain of definition of electromagnetic fields, and leads us to complete it as follows. For each value of $\pp$, we include not one, but two different branches for the fields: One branch corresponds to the dynamic fields with $\omega=\cz|\pp|$ in \Eq{eq:Hel}, and the other corresponds to static fields with $\omega=0$. When we complete \Eq{eq:Hel} in this way, we obtain the definition of the total integrated helicity: 
{\small
\begin{equation}
	\label{eq:Helbranches}
	\boxed{
	\begin{split}
		&\AVLambdatot =\AVLambdadynamic+\AVLambdastatic =\\
		&\intdpconf\Gpplusdynamicdagger\Help\Gpplusdynamic\\
		&+\Gpminusdynamicdagger\Help\Gpminusdynamic\\
		&+\intdpconf\Gpplusstaticdagger\Help\Gpplusstatic+\Gpminusstaticdagger\Help\Gpminusstatic.\\
	\end{split}
	}
\end{equation}
}

For $\AVLambdadynamic$, we can use the well-known fact that, since $\Help\Epdyn=\ii\cz\Bpdyn$, and $\Help\cz\Bpdyn=-\ii\Epdyn$, the $\Gpplusminusdynamic$ are eigenstates of the helicity operator with eigenvalue $\pm$1
\begin{equation}
	\Help\Gpplusminusdynamic=\pm\Gpplusminusdynamic,
\end{equation}
to rewrite $\AVLambdadynamic$ in \Eq{eq:Helbranches} 
\begin{equation}
	\label{eq:Helbranches2}
	\begin{split}
	&	\AVLambdatot =\AVLambdadynamic+\AVLambdastatic=\\
		&\intdpconf|\Gpplusdynamic|^2-|\Gpminusdynamic|^2+\\
&\intdpconf\Gpplusstaticdagger\Help\Gpplusstatic+\Gpminusstaticdagger\Help\Gpminusstatic.\\
	\end{split}
\end{equation}

We now set out to work on $\AVLambdastatic$ by elucidating the action of $\Help$ on $\Gpplusminusstatic$. To such end, we will use the wavevector-space version of \Eq{eq:staticeqs}:
\begin{equation}
	\label{eq:staticeqspp}
	\begin{split}
		\ii\pp\cdot\Ep=\frac{\rhorestpp}{\epsilon_0}&,\ \ii\pp\times\Ep=\zerovec,\\
		\ii\pp\cdot\Bp=\zerovec&,\ \ii\pp\times\Bp=\muz\ii\pp\times\mrestpp.\\
	\end{split}
\end{equation}
The longitudinal character of the $\mathbf{E}$ and the transverse character of $\mathbf{B}$ are manifest in Eqs.~(\ref{eq:staticeqs},\ref{eq:staticeqspp}): $\nabla\times\Er=\zerovec$, $\ii\pp\times\Ep=\zerovec$, $\nabla\cdot\Br=\zerovec$, $\ii\pp\cdot\Bp=0$. Therefore, when applying the helicity operator $\Help$ to $\Gpplusminusstatic$, the electric field $\Ep$ vanishes (see Tab.~\ref{tab:decomphel})
\begin{equation}
	\label{eq:rsstatic}
	\begin{split}
		\Help\Gplambda&=\Help\sqrt{\frac{\epsz}{2}}\left[\Ep\pm\ii \cz\Bp\right]\\&=\pm\ii\sqrt{\frac{1}{2\muz}}\Help\Bp,
	\end{split}
\end{equation}
and we see that the static $\Gplambda$ are not helicity eigenstates, in contrast to the dynamic case. Now, the purely transverse $\Bp$ in \Eq{eq:rsstatic} can be decomposed into two pieces of well-defined and opposite helicity $\lambda=\pm 1$, with an obvious action of $\Help$ on each of them (Tab.~\ref{tab:decomphel}):
\begin{equation}
	\label{eq:helpbp}
	\begin{split}
		\Bp&=\Bpplus+\Bpminus \implies \\
		\Help\Bp&=\Bpplus-\Bpminus,
	\end{split}
\end{equation}
with which \Eq{eq:rsstatic} changes into
\begin{equation}
	\label{eq:helstatic2}
	\Help\Gplambda=\pm\ii\sqrt{\frac{1}{2\muz}}\left[\Bpplus-\Bpminus\right].
\end{equation}

Using \Eq{eq:rsstatic}, \Eq{eq:helstatic2}, and Tab.~\ref{tab:decomphel} we can readily see that \footnote{\begin{equation*}
	\begin{split}
		&	\Gplambdadagger\Help\Gplambda\duetoeq{eq:rsstatic}\\
		&\sqrt{\frac{\epsz}{2}}\left[\Ep\pm\ii \cz\Bp\right]^\dagger(\pm\ii)\sqrt{\frac{1}{2\muz}}\Help\Bp\\
								&=\sqrt{\frac{\epsz}{2}}{\Ep}^\dagger(\pm\ii)\sqrt{\frac{1}{2\muz}}\Help\Bp\\+&\sqrt{\frac{\epsz}{2}}\cz{\Bp}^\dagger\sqrt{\frac{1}{2\muz}}\Help\Bp\\
								&\equaldueto{\text{Tab.~\ref{tab:decomphel}}}\sqrt{\frac{\epsz}{2}}\cz{\Bp}^\dagger\sqrt{\frac{1}{2\muz}}\Help\Bp\\
								&\equaldueto{\text{Tab.~\ref{tab:decomphel}}}\frac{1}{2\muz}\left[\Bpplus+\Bpminus\right]^\dagger\left[\Bpplus-\Bpminus\right]\\&\equaldueto{\text{Tab.~\ref{tab:decomphel}}}\frac{1}{2\muz}\left[|\Bpplus|^2-|\Bpminus|^2\right].
	\end{split}
\end{equation*}
}

\begin{equation}
	\label{eq:inthere}
	\begin{split}
		\Gplambdadagger&\Help\Gplambda\\
		&=\sqrt{\frac{\epsz}{2}}\cz{\Bp}^\dagger\sqrt{\frac{1}{2\muz}}\Help\Bp\\
		&=\frac{1}{2\muz}\left[|\Bpplus|^2-|\Bpminus|^2\right],\\
	\end{split}
\end{equation}
which we can substitute in \Eq{eq:Helbranches2} 
\begin{equation}
	\label{eq:Helbranches3}
	\boxed{
	\begin{split}
		\AVLambdatot =&\AVLambdadynamic+\AVLambdastatic=\\
		&\intdpconf|\Gpplusdynamic|^2-|\Gpminusdynamic|^2\\
&+\intdpconf \frac{|\Bpplus|^2-|\Bpminus|^2}{2\muz}.
	\end{split}
	}
\end{equation}
The newly discovered contribution of the static $\mathbf{B}$ field is added to the integrated value of the dynamic electromagnetic helicity. We will now show that $\AVLambdastatic$ is nothing but the magnetic helicity in different units. Let us use the second and third lines in \Eq{eq:inthere} to write
\begin{equation}
	\label{eq:twolast}
	\begin{split}
		\AVLambdastatic&=\intdpconf\frac{|\Bpplus|^2-|\Bpminus|^2}{2\muz}\\
		&=\intdpconf \frac{1}{2\muz}\Bpdagger\boxed{\frac{\Help}{|\pp|}\Bp},
	\end{split}
\end{equation}
and work on the expression inside the box. We consider the relationship between the magnetic field and the magnetic vector potential $\Br=\nabla\times\Ar$ in $\pp$-domain: $\Bp=\ii\pp\times \Ap$, and operate on both its sides with $\frac{\Help}{|\pp|}$ from the left:
\begin{equation}
	\label{eq:onelast}
	\begin{split}
		&	\frac{\Help}{|\pp|}\Bp=\frac{\Help}{|\pp|}\ii\pp\times\Ap=(\Help)^2\Ap\equaldueto{\text{Tab.~\ref{tab:decomphel}}}\\&\Aptrans=\Ap-\mathbf{A}^\parallel(\pp)=\Ap-\phat\left[\phat\cdot\Ap\right].
	\end{split}
\end{equation}
We substitute the last expression in \Eq{eq:onelast} into the box in \Eq{eq:twolast} to get:
\begin{equation}
	\begin{split}
		\AVLambdastatic&=\int_{\mathbb{R}^3} \frac{d \pp}{2\Zz}\text{ } \Bpdagger\left\{\Ap-\phat\left[\phat\cdot\Ap\right]\right\}\\
		&=\int_{\mathbb{R}^3}  \frac{d \pp}{2\Zz}\text{ } \Bpdagger\Ap,
	\end{split}
\end{equation}
where the second equality follows because the longitudinal $\mathbf{A}^\parallel(\pp)$ is canceled by the projection with the transverse $\Bp$ (Tab.~\ref{tab:decomphel}). After using Parseval's theorem (see e.g. \cite[Eq.~B3, I.B.1]{Cohen1997}) we reach 
\begin{equation}
	\label{eq:avlmh} 
		\AVLambdastatic=\int_{\mathbb{R}^3} \frac{d \rr}{2\Zz}\text{ } \Br\cdot\Ar,
\end{equation}
which is, essentially, the magnetic helicity of the static magnetic field \cite{Woltjer1958,Moffatt1969,Ranada1992}. The differences with the typical definition are a factor of 1/2, and a factor of $1/\Zz$ that endows $\AVLambdastatic$ with units of angular momentum, matching the units of electromagnetic helicity. It is important to note that the cancellation of the longitudinal part of $\Ap$ due to the transverse character of $\Bp$ happens independently of the chosen gauge. Therefore, the derivation leading to \Eq{eq:avlmh} is gauge independent. It is also important to note that $\AVLambdastatic$ is invariant under duality transformations. The lack of this invariance has been used in Ref.~\onlinecite{Barnett2012} as an argument for considering the magnetic helicity to be essentially different from the electromagnetic helicity of the free field. The key point for establishing the invariance is the action of the helicity operator on the static $\Ep$ and $\Bp$ fields
\begin{equation}
	\label{eq:goback}
	\begin{split}
		\Help\Ep&\duetoeq{eq:staticeqspp}\zerovec,\\
		\Help\Bp&\duetoeq{eq:helpbp}\Bpplus-\Bpminus,
	\end{split}
\end{equation}
which shows that, {\em in the static case the electromagnetic duality transformation does not mix the electric and magnetic fields}, as it does in the dynamic case [\Eq{eq:d}]. It rather has the following effects, which are readily derived from \Eq{eq:goback} and the construction of the duality transformation as the exponentiation of the helicity operator $D_\theta=\exp\left(-\ii\theta\Lambda\right)$:
\begin{equation}
	\label{eq:aipobret}
	\begin{split}
		D_\theta\Ep&=\exp(-i\theta\Help)\Ep=\Ep,\\
		D_\theta\Bpplus&=\exp(-i\theta\Help)\Bpplus=\Bpplus\exp(-\ii\theta),\\
		D_\theta\Bpminus&=\exp(-i\theta\Help)\Bpminus=\Bpminus\exp(\ii\theta).
	\end{split}
\end{equation}
The duality invariance of $\AVLambdastatic$ can be seen by substituting the last two lines of \Eq{eq:aipobret} into the first line of \Eq{eq:twolast}. This, together with the well-known invariance of $\AVLambdadynamic$ under duality, implies the invariance of the total helicity $\AVLambdatot$ in \Eq{eq:Helbranches3}.

Let us now express $\AVLambdastatic$ as a function of the magnetization density. It readily follows from \Eq{eq:staticeqspp} that $\Bp=\muz\mresttranspp$, showing that helicity can be stored in the transverse part of the magnetization. We can rewrite \Eq{eq:Helbranches3} as
\begin{equation}
	\label{eq:Helbranches4}
	\boxed{
	\begin{split}
		\AVLambdatot =&\intdpconf|\Gpplusdynamic|^2-|\Gpminusdynamic|^2
		\\&+\intdpconf \frac{|\mrestppplus|^2-|\mrestppminus|^2}{2/\muz}
.
	\end{split}
	}
\end{equation}

The definition of the total helicity in Eqs.~(\ref{eq:Helbranches2},\ref{eq:Helbranches3},\ref{eq:Helbranches4}) unifies the static magnetic helicity and the dynamic electromagnetic helicity into a single quantity. This total helicity is the sum of two terms that measure the difference between the number of left-handed and right-handed photons of the free field, and the screwiness of the static magnetization, respectively. While the magnetic and electromagnetic helicities have previously been discussed together \cite{Ranada1992,Afanasiev1996,Trueba1996,Nienhuis2016}, they have, as far as I know, not been unified into a single physical property until now. According to this unification, the static and dynamic helicities are two manifestations of the same fundamental quantity in different frequency regimes. As such, they are susceptible to change into each other, giving a positive answer to the question of whether helicity can be exchanged between light and matter. Regarding the question of the effects of such exchange: Equation~(\ref{eq:Helbranches4}) indicates that systems able to sustain static magnetization configurations with some degree of screwiness have the potential for storing electromagnetic helicity coming from the free dynamic field. Such storage implies a modification of the transverse part of the initial static magnetization. Conversely, such systems have the potential for returning the stored helicity to the free field by means of electromagnetic radiation, with a corresponding change in their static magnetization state.

\section{Conclusions and discussion\label{sec:conclusion}}
In conclusion, this article shows that the electromagnetic helicity of the free electromagnetic field, and the magnetic helicity of the static magnetization are two different parts of the same physical quantity, the total helicity. The total helicity is the sum of two terms. One term quantifies the screwiness of the static magnetization in matter, and the other quantifies the difference between the number of left- and right- handed photons in the free electromagnetic field. The unification establishes the theoretical basis for studying the conversion between these two embodiments of helicity upon light-matter interaction. 

Both manifestations of helicity are separately relevant in quite diverse areas of physics. Electromagnetic helicity is particularly relevant in chiral light-matter interactions, and magnetic helicity is relevant in areas like cosmology \cite{Vachaspati2001,Caprini2004}, solar physics \cite{Berger1999}, fusion physics \cite{Jarboe1983,Jarboe1989}, magneto-hydrodynamics \cite{Hirono2015,Avdoshkin2016,Mace2020,Figueroa2019}, and condensed matter \cite{Beaurepaire1996,Stanciu2007,Nagaosa2013,Garst2017,Kaushik2019}. Consequently, their unified understanding has the potential for simultaneously impacting several different fields. For example, the new link between optics and magnetism is apparently relevant for the physics of all optical switching of magnetization with circularly polarized radiation \cite{Stanciu2007}, and for the optical control of helical magnets and skyrmions \cite{Muehlbauer2009,Nagaosa2013,Garst2017}. I believe that other areas where the results of this paper could be useful are fusion physics, where the injection of magnetic helicity is considered for controlling the plasma \cite{Jarboe1983,Jarboe1989}, and cosmology, where helical magnetic fields with galactic-scale coherent lengths are studied\cite{Vachaspati2001,Caprini2004}.
\begin{acknowledgments}
	This work was funded by the Deutsche Forschungsgemeinschaft (DFG, German Research Foundation) -- Project-ID 258734477 -- SFB 1173.
\end{acknowledgments}

\end{document}